\documentclass[%
 reprint,prc,
amsmath,amssymb,
aps,
floatfix,
]{revtex4-1}

\usepackage{graphicx}% Include figure files
\usepackage{dcolumn}% Align table columns on decimal point
\usepackage{bm}% bold math
\usepackage{url}
\usepackage[colorlinks=true,linktocpage=true,linkcolor=blue,citecolor=blue,
allcolors=blue] {hyperref }
\usepackage[mathlines]{lineno}% Enable numbering of text and display math
\usepackage{times} 
\usepackage{lipsum}% http://ctan.org/pkg/lipsum

\begin{document}
\preprint{APS/123-QED}

\title{Rotation modified QCD equation of state across crossover at finite chemical potential}
\author{S. Ipsita Sahoo$^{1,2}$}%
\email{sisahoo@barc.gov.in}
\author{Sanjeebani Biswal$^{2}$}
\email{sanjeebani@barc.gov.in}
\author{Dipanwita Dutta$^{1,2}$}
\email{ddutta08@hbni.ac.in}
\author{Dipak Kumar Mishra$^{1,2}$}
\email{dkmishra@barc.gov.in}

\affiliation{$^1$Bhabha Atomic Research Centre, Mumbai - 400085, INDIA}
\affiliation{$^2$Homi Bhabha National Institute, Anushaktinagar, Mumbai - 400094, INDIA}

\date{\today}% It is always \today, today,

\begin{abstract}
It is well-established that at low baryon chemical potential, the transition 
from the confined hadronic phase to the deconfined partonic phase is a smooth 
crossover rather than a first or second-order phase transition.
We develop a thermodynamically consistent hybrid equation of state to study
the effect of global rotation on the equation of state across the QCD
crossover that interpolates the hadronic and partonic phases.
The temperature dependence of normalized thermodynamic observables, such as
entropy density ($s/T^3$), pressure ($P/T^4$), energy density ($\varepsilon/T^4$),
specific heat ($C_V/T^3$), and the square of speed of sound ($c_s^2$) 
in presence of rotation are investigated. The effect of
non-zero chemical potential on thermodynamic quantities in presence of
rotation is also studied. Our results show that increasing chemical
potential enhances the thermodynamic quantities, whereas rotation
suppresses them in the crossover region. Further, we investigate
the effect of rotation on the conserved numbers susceptibilities
and their correlations as a function of temperature. The susceptibilities
and their correlations exhibit a systematic enhancement with increase
in rotation in both the hadronic and partonic phases. These findings
provide new insights into the interplay between rotation and QCD
thermodynamics, which can have important implications for rapidly
rotating strongly interacting matter produced in ultra-relativistic
heavy-ion collisions.
\end{abstract}

%\pacs{24.85.+p,25.30.Dh,25.75.-q}% PACS, the Physics and Astronomy
\maketitle

\section{Introduction}\label{intro}
The study of strongly interacting matter under extreme conditions of
temperature and energy density remains one of the central objectives of
research programs at the Relativistic Heavy-Ion
Collider (RHIC) and the Large Hadron Collider (LHC). According to the
theory of Quantum Chromodynamics (QCD), under such extreme conditions,
an ordinary hadronic matter undergoes a transition from confined state
to a deconfined state of
quarks and gluons, known as the quark-gluon plasma
(QGP)~\cite{Bzdak:2019pkr,Rischke:2003mt,Aoki:2006br,Borsanyi:2015waa,Aoki:2006we}.
First principle lattice QCD calculations have established that,
at vanishing or small
baryon chemical potential, the transition from the confined hadronic phase
(HG) to the deconfined QGP phase is not a simple first or second-order phase
transition but rather a smooth
crossover~\cite{Aoki:2006we,Bowman:2008kc}. Instead of exhibiting discontinuities 
or latent heat, the crossover is characterized by rapid and continuous
changes in thermodynamic observables, such as the pressure, energy density,
entropy density, and susceptibilities of the conserved quantities
such as baryon number, electric charge and strangeness.
On the other hand, at larger baryon chemical potential the system may
experience a first-order phase transition terminating at the
critical end point (CEP)~\cite{Bazavov:2019lgz, Ejiri:2008xt, Bowman:2008kc}. 
Consequently, developing theoretical models capable of describing both confined 
hadronic and deconfined QGP phase while reproducing the crossover has become an
important theoretical challenge.

The thermodynamics of the hadronic phase below the critical temperature
($T_c\sim 156$ MeV) is successfully described
by the hadron resonance gas (HRG) model, which treats hadrons and
resonances as effective degrees of freedom~\cite{Andronic:2012ut,Karsch:2011,Garg:2013}. 
The HRG model reproduces lattice-QCD calculations and provides an excellent 
description of hadron yields observed in heavy-ion collisions below 
$T_c$~\cite{HotQCD:2014kol, Bollweg:2021vqf, Tawfik:2004sw, Braun-Munzinger:2003pwq}. 
Nevertheless, the ideal HRG model becomes inadequate near the QCD transition due to
neglected hadronic interactions, finite-size effects. 
Several extensions of the HRG model have been proposed to incorporate these
effects and develop a unified equation of state that smoothly
connects the hadronic and partonic phases~\cite{Yang:2026brr, Miyahara:2019zfn}.

Above the critical temperature, the relevant degrees of freedom are quarks
and gluons, and the QGP is commonly described using perturbative QCD,
quasi-particle approaches, or effective field 
theories~\cite{Mykhaylova:2020pfk,Thaler:2003uz,Sambataro:2024mkr}.
However, these models must be smoothly connected to the hadronic phase
near the transition. Abrupt
matching between HG and QGP descriptions can lead to thermodynamic
inconsistencies, affecting the accuracy of predictions for
phase-transition observables. A realistic description of QCD matter
requires a hybrid framework that unifies hadronic and partonic phases
while preserving thermodynamic continuity over a broad range of
temperatures and baryon chemical 
potentials~\cite{Qin:2025xvw,Asakawa:1995zu,Qin:2026hub}.

Recent experimental and theoretical developments have further highlighted
the importance of rotational effects in strongly interacting 
matter~\cite{Jiang:2016wvv, Fujimoto:2021xix, STAR:2017ckg}.
Measurements of global polarization by the STAR Collaboration have
demonstrated that non-central heavy-ion collisions produce the most
inviscid fluid ever observed, with vorticities reaching approximately
$10^{21}$~s$^{-1}$~\cite{STAR:2017ckg}. In non-central collisions, 
the resulting fireball can sustain rapid rotation, generating a large 
angular momentum that induces vorticity in the 
fluid~\cite{Deng:2016gyh, Jiang:2016woz}.
Such extreme vorticity can substantially modify the thermodynamic
properties of QCD matter, alter the equation of state, and influence
the structure of the QCD phase 
diagram~\cite{Fujimoto:2021xix, Mukherjee:2023ijv, Wang:2018sur, Wei:2021dib}. 
Understanding the role of rotation has therefore emerged as an important frontier 
in heavy-ion physics.

In addition to the thermodynamic observables, fluctuations and
correlations of the conserved charges are sensitive to the effect of 
rotation~\cite{Mukherjee:2023ijv, Padhan:2026mwg}.
These quantities exhibit distinct behavior in the hadronic and the QGP
phase, and are expected to be largely enhanced in the vicinity of the
critical end point~\cite{Asakawa:2000wh, Jeon:2000wg, Luo:2017faz, Stephanov:2011pb}. 
Several recent studies have investigated the influence
of rotation separately within hadronic and partonic descriptions of QCD
matter. The effects of rotation have been included in HRG to study the EOS 
of QCD matter~\cite{Fujimoto:2021xix}. The EOS was further extended to study 
the effect of rotation in the QGP phase. However, a systematic analysis of 
rotational effects within a unified crossover equation of state remains 
largely unexplored.

Motivated by these considerations, we employ a hybrid hadron gas
(HG)--quark--gluon plasma (QGP) framework in which an interacting
hadronic equation of state is smoothly connected to the deconfined
phase through a thermodynamically consistent crossover construction, 
in the presence of rotation.
Such a framework provides a unified description of hadronic and
partonic matter under rotation, reproducing key thermodynamic observables 
while ensuring smooth behavior across the QCD transition, making it well
suited for studies of heavy-ion collisions and QCD transition 
phenomena~\cite{Asakawa:1995zu}.

In the present work, we investigate the influence of global rotation on the
thermodynamic properties of QCD matter within a unified hybrid HG--QGP
framework~\cite{Asakawa:1995zu,Qin:2026hub}.
%that smoothly interpolates between the hadronic and partonic phases
We examine the temperature
dependence of the thermodynamic
quantities and equation of state in the presence of finite angular
velocity and baryon chemical potential. Furthermore, we study the
rotational dependence of conserved-charge susceptibilities and their
correlations across the crossover region.

The paper is organized as follows. In section~\ref{rotw} and~\ref{qgp}, 
we discuss the formalism of modified HRG model including rotation, and
construct the EOS of the  QCD matter in the hadronic and QGP phase in
presence of global rotation.
In section~\ref{cross}, we describe the thermodynamically consistent
interpolation employed to construct a smooth crossover between the
hadronic and partonic equations of state.
The formalism is further extended to study the effect of rotation on quadratic 
fluctuations of conserved numbers and their correlations in section~\ref{chi}.
We discuss the results and their implications in section~\ref{results}.
Finally, the conclusions and perspectives are summarized in
Sec.~\ref{conclusion}.

\section{HRG model with rotation}\label{rotw}
In the presence of global rotation, the single-particle energy dispersion
relation is modified due to the coupling between the particle's angular
momentum and the rotational field~\cite{Fujimoto:2021xix}. Consequently, rotation
introduces an effective chemical potential, leading to the modified
energy dispersion relation given by
\begin{equation}
    E_{l,i}=\sqrt{k_r^2+k_z^2+m_i^2}-(l+S_i)\omega.
\end{equation}
$S_i$ and $m_i$ are the spin and mass of the $i^{th}$ particle. 
The generalized pressure for the system of baryons and mesons in the presence 
of rotation is given by the sum total of pressure of each particle, $i$, as 
below:
\begin{equation}
    P^{HRG} = \sum_i P_i^{b/m},
\end{equation}
where
\begin{equation}\label{ph}
\begin{split}
    P_i^{b/m} = \pm \frac{T}{8\pi^2} \sum_{l=-\infty}^{\infty} 
    \int_{(\Lambda_l^{IR})^2} dk_r^2 \int dk_z \sum_{\nu=l}^{l+2S_i} 
J_{\nu}^2(k_rr) \\ \times \log \left[1\pm 
\exp\left(-\frac{E_{l,i}-\mu_i}{T}\right)\right].
\end{split}
\end{equation}
We note that an infrared cutoff for the $k_r$ integration, $\Lambda_l^{IR} 
= \zeta_{l,1}\omega$, is introduced. Here, $\zeta_{l,1}$ is the first zero of 
the Bessel function ($J_l(\zeta_{l,1})=0$). The cutoff arises from the causality 
bound and guarantees the relation $E_{l,1} \ge 0$.

The entropy density of the hadronic system is obtained from 
Eq.~\ref{ph} and the relation $s = \partial P/\partial T$ as:
\begin{equation}
    s^{HRG}= \sum_i s_i^{b/m},
\end{equation}
where
\begin{equation}
\begin{split}
    s_i^{b/m} =  \pm \frac{1}{8\pi^2} \sum_{l=-\infty}^{\infty} 
    \int_{(\Lambda_l^{IR})^2} dk_r^2 \int dk_z \sum_{\nu=l}^{l+2S_i} 
J_{\nu}^2(k_rr) \\ \times \Bigg\{\log \left[1\pm 
\exp\left(-\frac{E_{l,i}-\mu_i}{T}\right)\right] \\
    \pm \left(\frac{E_{l,i}-\mu_i}{T}\right)\frac{1}
    {\exp\left(\frac{E_{l,i}-\mu_i}{T}\right)\pm1}\Bigg\}.
\end{split}
\end{equation}

\section{Rotating quark-gluon gas}\label{qgp}
The thermodynamics of QGP can be described by quark and gluon degrees of 
freedom. The EOS of this matter is of fundamental importance to understand the 
properties of this system.
The gluon pressure is:
%\begin{widetext}
\begin{equation}\label{pg}
\begin{split}
    P_g = -\frac{T}{8\pi^2} \sum_{l=-\infty}^{\infty} \int_{(\Lambda_l^{IR})^2} 
    dk_r^2 \int dk_z \left[J_l^2(k_rr) + J_{l+2}^2(k_rr)\right]\\
    \times \log 
\left[1 - \exp\left(-\frac{\sqrt{k_r^2+k_z^2}-(l+1)\omega}{T}\right)\right].
\end{split}
\end{equation}
%\end{widetext}
Since gluons are massless gauge bosons, there is no contribution from $s_z=0$. 
Hence, the term $J_{l+1}^2(k_rr)$ does not appear in the above equation.

From Eq.~\ref{pg}, the entropy density for gluon is obtained as:
\begin{equation}
\resizebox{\columnwidth}{!}{$
    \begin{split}
        s_g = -\frac{1}{8\pi^2} \sum_{l=-\infty}^{\infty} 
        \int_{(\Lambda_l^{IR})^2} dk_r^2 \int dk_z \left[J_l^2(k_rr) + J_{l+2}^2(k_rr)\right] \\
        \times \Bigg\{\log \left[1 -\exp\left(-\frac{\sqrt{k_r^2+k_z^2}-(l+1)\omega}{T}\right)\right] \\
        - \left(\frac{\sqrt{k_r^2+k_z^2}-(l+1)\omega}{T}\right)        
\frac{1}{\exp\left(\frac{\sqrt{k_r^2+k_z^2}-(l+1)\omega}{T}\right)-1}\Bigg\}.
    \end{split}
    $}
\end{equation}

The quark (anti-quark) pressure is:
\begin{equation}\label{pq}
\begin{split}
    P_i^{q/\bar q} = \frac{T}{8\pi^2} \sum_{l=-\infty}^{\infty} 
    \int_{(\Lambda_l^{IR})^2} dk_r^2 \int dk_z \left[J_l^2(k_rr) + 
      J_{l+1}^2(k_rr)\right]\\
    \times \log \left[1 +
\exp\left(-\frac{\sqrt{k_r^2+k_z^2+m_i^2}-(l+\frac{1}{2})\omega \mp 
\mu_i}{T}\right)\right].
\end{split}
\end{equation}
From Eq.~\ref{pq}, the entropy density for quark (anti-quark) is obtained as:
\begin{equation}
%\resizebox{\columnwidth}{!}{$
    \begin{split}
        s_i^{q/\bar q} = \frac{1}{8\pi^2} \sum_{l=-\infty}^{\infty}
        \int_{(\Lambda_l^{IR})^2} dk_r^2 \int dk_z \left[J_l^2(k_rr) + 
J_{l+1}^2(k_rr)\right]\\
        \times \Bigg\{\log \left[1 + \exp\left(-\frac{\sqrt{k_r^2+k_z^2+m_i^2}-
        (l+\frac{1}{2})\omega \mp \mu_i}{T}\right)\right] \\
        + \left(\frac{\sqrt{k_r^2+k_z^2+m_i^2}-(l+\frac{1}{2})\omega \mp \mu_i}{T}\right)\\
        \times \frac{1}{\exp\left(\frac{\sqrt{k_r^2+k_z^2+m_i^2}-(l+\frac{1}{2})\omega \mp \mu_i}{T}\right)+1}\Bigg\}.
    \end{split}
%    $}
\end{equation}
From the normalization given by the Stefan-Boltzmann (SB) limit of a rotating 
quark-gluon gas, we have:
\begin{equation}\label{eflavor}
    P_{SB}\equiv (N_c^2-1)P_g + N_cN_f(P_q+P_{\bar q}),
\end{equation}
where, $N_c$ and $N_f$ are the number of colors and flavors respectively.
Since $N_c=3$, the degeneracy for gluons is $8$. 

\section{Smooth Crossover}\label{cross}
Lattice-QCD calculations with dynamical quarks reveal several valuable
insights into the thermodynamic properties of strongly interacting matter 
above the critical temperature, $T_c$. At $T_c$, there are several 
characteristic features such as: (a) there is a sharp 
increase in normalized entropy density, $s/T^3$, indicating the transition
from hadronic phase to QGP phase, (b) the normalized
pressure, $P/T^4$, gradually approaches the Stefan–Boltzmann limit
with increasing temperature, indicating the progressive liberation
of partonic degrees of freedom, (c) the scaled energy
density, $\varepsilon/T^4$, exhibits a pronounced peak just above $T_c$
before decreasing toward its asymptotic high-temperature value, and
(d) the interaction measure, $\varepsilon-3P$, remains finite above $T_c$,
indicating significant deviation from ideal-gas limit~\cite{Asakawa:1995zu}.

Among the various thermodynamic observables, the entropy density provides
a particularly suitable quantity for constructing a unified equation
of state. Once the entropy density is specified, all other bulk thermodynamic
observables such as pressure, energy density,
specific heat, and the squared speed of sound can be determined in a
self-consistent manner.
In the present framework, we employ the entropy density as the fundamental
interpolation variable to construct a smooth crossover between the
hadronic and partonic equations of state. Other quantities are
conveniently expressed in terms of the entropy density $(s(T))$ as
follows~\cite{Qin:2026hub}:

\begin{gather} 
P(T) = \int_{0}^{T} s(t)dt, \label{equP} \\ 
\varepsilon (T) = Ts(T) - P(T) + \sum_i \mu_i \int_0^T
\frac{\partial s(t)}{\partial \mu_i}dt, \label{equE} \\
C_v(T) = \left.\frac{\partial \varepsilon(T)}{\partial T}\right|_{V} 
= T\frac{\partial s(T)}{\partial T} + \sum_i \mu_i \frac{\partial s(T)}{\partial 
\mu_i}, \label{equCv} \\
c_s^2 = \frac{\partial P(T)}{\partial \varepsilon(T)} 
= \frac{s(T)}{C_V(T)}. \label{equCsq}
\end{gather}

\begin{figure}[hbp]
    \centering
    \includegraphics[width=\linewidth]{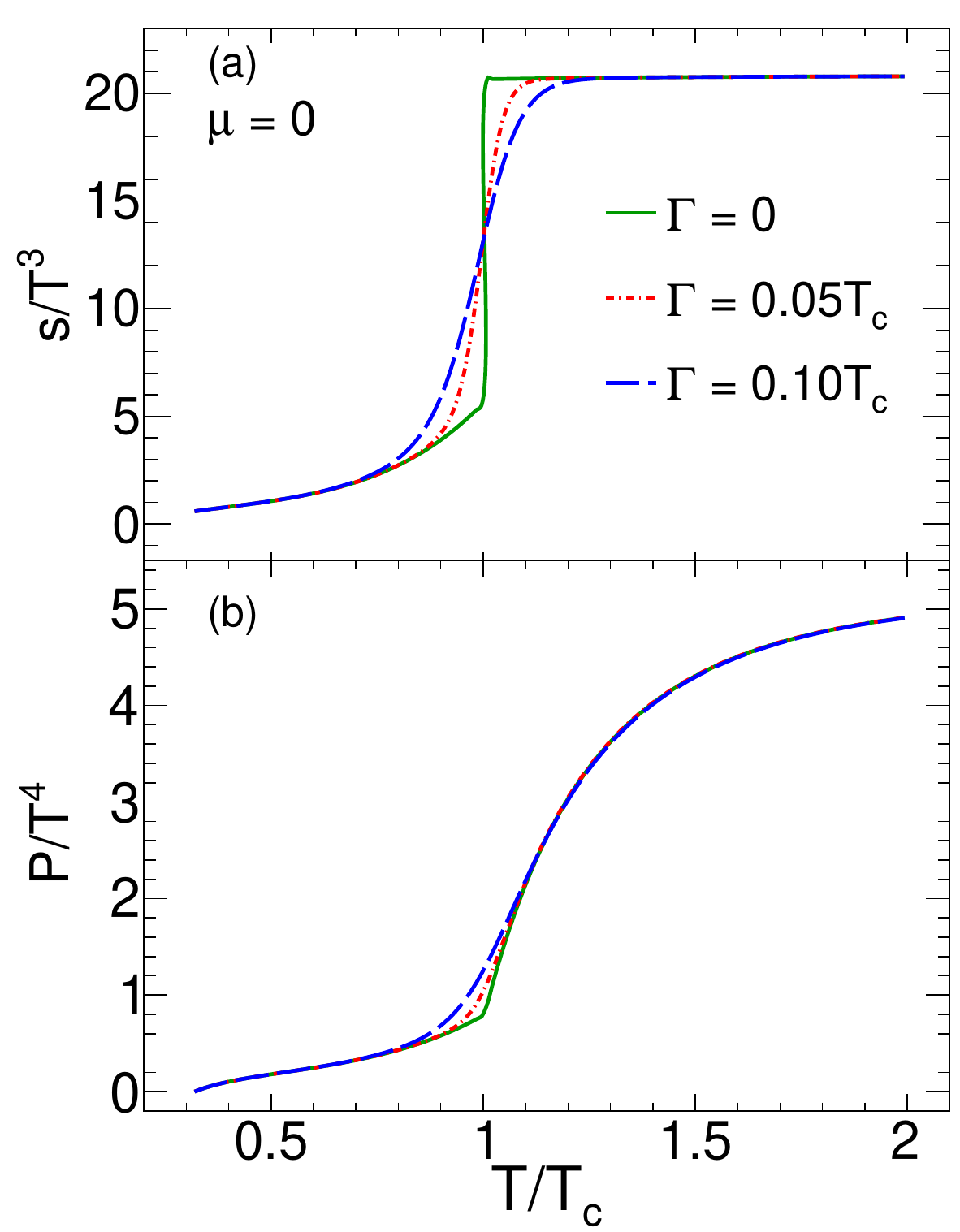}
    \caption{The temperature dependence of normalized entropy density and 
    normalized pressure at different $\Gamma$ values ($0$, $0.05Tc$, $0.10Tc$) 
    at $\mu=0$ and $\omega=0$.}
    \label{fig1}
\end{figure}

%From this normalization, we have $P(T=0)=0$ and $\varepsilon(T=0)=0$,
%which is also taken in the case of lattice calculation.
%We are interested in their gross behavior at the transition, and not the
%order of the transition.
In order to achieve a thermodynamically consistent transition between
the hadronic and partonic phases, we construct a smooth
interpolation of entropy density, $s(T)$, connecting hadron resonance
gas at low temperature and the quark-gluon plasma at high
temperature~\cite{Asakawa:1995zu}.
The entropy density, $s(T)$, is constrained by the thermodynamic inequality and 
Nernst's theorem such that:
\begin{equation} \label{Sineq}
\begin{aligned}
\frac{\partial s(T)}{\partial T} \geq 0, \mathrm{and}~~s(T)\rightarrow 
\mathrm{constant ~as~~} T\rightarrow 0
\end{aligned}
\end{equation}

The simplest parameterization of $s(T)$ that ensures continuity across the
crossover region while satisfying the fundamental thermodynamic consistent
condition in Eq.~\ref{Sineq} is:
\begin{equation}\label{eq:s}
    s(T) = f(T) s^{HRG}(T) + \left\{1-f(T)\right\} s^{QGP}(T).
\end{equation}
Here, $s^{HRG}(T)$ and $s^{QGP}(T)$ are the entropy densities, respectively, 
in hadronic and QGP phase including the effect of rotation as discussed in 
previous sections, and $f(T)$, the weighting function, is defined as~\cite{Qin:2026hub}:
\begin{equation}\label{gammafunc}
    f(T) = \frac{1}{2} \left[1-\mathrm{tanh}\left(\frac{T-T_c}{\Gamma}\right)
    \right].
\end{equation}

In Eq.~\ref{gammafunc}, we see that $f(T)$ is a decreasing function of 
temperature and $\Gamma$ sets the width of the transition region. 
In the region where $T$ satisfies $|T-T_c| > \Gamma$, $s(T)$ approaches 
$s^{HRG}(T)$ below $T_c$, i.e., the hadronic phase and $s^{QGP}(T)$ above
$T_c$, i.e., the QGP phase.
This is consistent with the assumption for crossover nature of the
QCD transition. The region where $T$ satisfies $|T-T_c| \le \Gamma$ is
interpreted as the crossover region. We can treat the crossover by changing
$\Gamma$ in a thermodynamic consistent way. Figure~\ref{fig1} shows $s(T)/T^3$
and $P(T)/T^4$ as a function of $T/T_c$ for $\Gamma/T_c = 0.0$, $0.05$, and
$0.10$. The entropy density at $\Gamma/T_c=0.05$ is similar to the lattice
measurements of $s(T)$, which shows a rapid variation 
across a highly constrained temperature interval of approximately $10$ MeV~\cite{Asakawa:1995zu}. 
In this study, $\Gamma$ is fixed as $0.05T_c$ and $T_c$ for phase transition is taken to be $156.5$ 
MeV~\cite{HotQCD:2018pds}. In Fig.~\ref{fig1} (a), for $\Gamma$ = 0, 
there is sharp change in the entropy
density, and as we increase $\Gamma$, $s(T)$ gets smoothen out. In
Fig.~\ref{fig1} (b), $P/T^4$ increases slowly above $T_c$ as compared 
to $s(T)$. 
Since $P(T)$ is calculated as an integral of $s(T)$, it is expected to observe such a
continuous and slow rise. Both $s/T^3$ and $P/T^4$ increase smoothly with
temperature, on increasing $\Gamma$.  

\section{Conserved number fluctuations and their correlations}\label{chi}

The ﬂuctuations and correlations of conserved charges, baryon number ($B$),
electric charge ($Q$), and strangeness ($S$), serve as sensitive probes
of thermodynamic properties and phase structure of strongly interacting
matter~\cite{Asakawa:2000wh,Jeon:2000wg,Ejiri:2005wq}.

They provide direct information on the underlying degrees of freedom 
and the relative contributions of different particle species to the 
equation of state across the QCD phase transition.
The fluctuations are characterized by the
corresponding susceptibilities of the conserved charges, which are defined as
the derivative of scaled pressure with respect to scaled chemical
potential~\cite{Bazavov:2013dta}:

\begin{equation}
    \chi_{kmn}^{BQS}=\frac{\partial^{k+m+n}(P/T^4)}{\partial(\mu_B/T)^k 
    \partial(\mu_Q/T)^m \partial(\mu_S/T)^n}.
\end{equation}
Here, $k$, $m$, and $n$ denotes the order of the derivatives with respect to
baryon, electric charge, and strangeness chemical potential. In this study,
we focus on the the quadratic fluctuations of conserved charges along with their
correlations in a rotating hadron gas which are expressed, respectively,
as below:
%\begin{widetext}
\begin{equation}\label{eq:chi2}
    \begin{split}
        \chi_2^X= \frac{X^2}{8\pi^2T^3} \sum_i \sum_{l=-\infty}^{\infty} 
        \int_{(\Lambda_l^{IR})^2} dk_r^2 \int dk_z \sum_{\nu=l}^{l+2S_i} 
        J_{\nu}^2(k_rr)\\
        \times \exp\left(\frac{E_{l,i}-\mu_i}{T}\right) 
\frac{1}{\left(\exp\left( \frac{E_{l,i}-\mu_i}{T}\right) \pm 1\right)^2}
    \end{split}
\end{equation}

\begin{equation}\label{eq:xy}
    \begin{split}
        \chi_{11}^{XY}= \frac{XY}{8\pi^2T^3} \sum_i \sum_{l=-\infty}^{\infty}
        \int_{(\Lambda_l^{IR})^2} dk_r^2 \int dk_z \sum_{\nu=l}^{l+2S_i} 
        J_{\nu}^2(k_rr)\\
        \times \exp\left(\frac{E_{l,i}-\mu_i}{T}\right) 
\frac{1}{\left(\exp\left( \frac{E_{l,i}-\mu_i}{T}\right) \pm 1\right)^2}
    \end{split}
\end{equation}
%\end{widetext}
In case of a rotating quark-gluon gas, the conserved number susceptibilities
and their correlations, respectively, are:
%\begin{widetext}
\begin{equation}\label{eq:chi2_rot}
 \resizebox{\columnwidth}{!}{$
    \begin{split}
        \chi_2^X= \frac{X^2}{8\pi^2T^3} \sum_i \sum_{l=-\infty}^{\infty} 
        \int_{(\Lambda_l^{IR})^2} dk_r^2 \int dk_z \left(J_l^2(k_rr) + 
        J_{l+1}^2(k_rr)\right)\\
        \times 
\exp\left(\frac{\sqrt{k_r^2+k_z^2+m_i^2}-(l+\frac{1}{2})\omega \mp 
\mu_i}{T}\right) \\
\frac{1}{\left(\exp\left(\frac{\sqrt{k_r^2+k_z^2+m_i^2}-(l+\frac{1}{2})\omega 
\mp \mu_i}{T}\right) + 1\right)^2}.
    \end{split}
    $}
\end{equation}

\begin{equation}\label{eq:xy_rot}
\resizebox{\columnwidth}{!}{$
    \begin{split}
        \chi_{11}^{XY}= \frac{XY}{8\pi^2T^3} \sum_i \sum_{l=-\infty}^{\infty} 
        \int_{(\Lambda_l^{IR})^2} dk_r^2 \int dk_z \left(J_l^2(k_rr) + 
        J_{l+1}^2(k_rr)\right)\\
        \times 
\exp\left(\frac{\sqrt{k_r^2+k_z^2+m_i^2}-(l+\frac{1}{2})\omega \mp 
\mu_i}{T}\right) \\
\frac{1}{\left(\exp\left(\frac{\sqrt{k_r^2+k_z^2+m_i^2}-(l+\frac{1}{2})\omega 
\mp \mu_i}{T}\right) + 1\right)^2}.
    \end{split}
    $}
\end{equation}
%\end{widetext}

In the above Eqs.~\ref{eq:chi2}, \ref{eq:chi2_rot}, $X$ denotes the conserved 
numbers $B$, $Q$, or $S$ and in Eqs.~\ref{eq:xy}, \ref{eq:xy_rot}, $XY$ denotes 
$BQ$, $QS$, or $BS$.

\section{Results and Discussions}\label{results}
\begin{figure}[htbp]
    \centering
    \includegraphics[width=\linewidth]{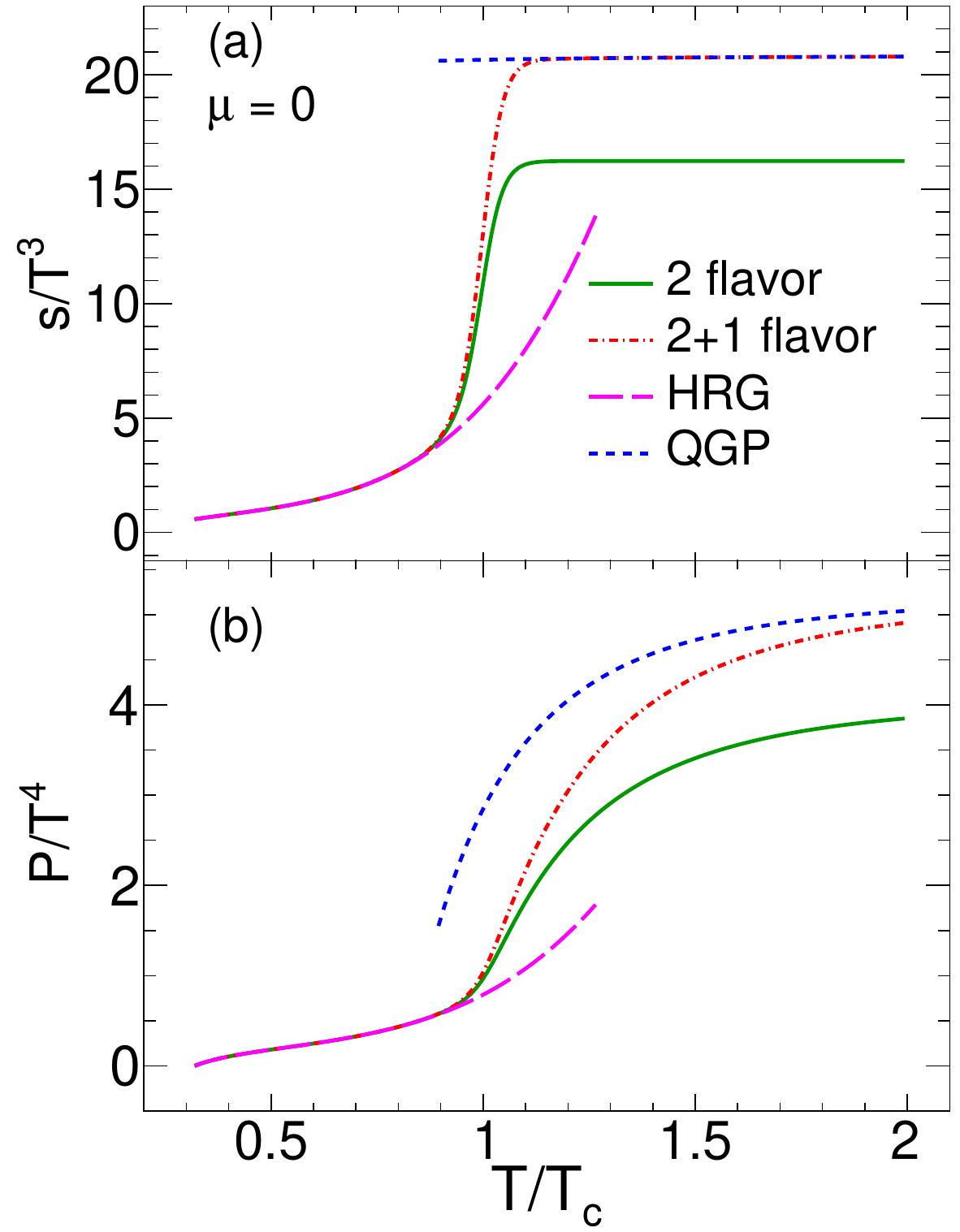}
    \caption{$s/T^3$ and $P/T^4$ as a function of $T/T_c$ at $\mu=0$ and $\omega=0$. 
      The green solid line correspond to $2$ flavor QGP state considering only the 
      light quarks ($u$ and $d$), while the red dashed line corresponds to the 
      $(2+1)$ flavor QGP state constituting 3 quarks with mass ($u$, $d$, and $s$). 
      The magenta line extrapolates the thermodynamic quantities in HRG beyond $T_c$, 
      while the blue line corresponds to the contribution from partonic phase only.}
    \label{fig2}
\end{figure}

Figure~\ref{fig2} shows the temperature dependence of the normalized
entropy density, $s/T^3$ (top panel) and normalized pressure $P/T^4$
(bottom panel) for both 2-flavor ($u$, $d$) and (2+1) flavor
($u$, $d$, $s$) QGP equation of state. In the hadronic sector at $T<T_c$,
the calculations are performed within the HRG model by including all the
hadrons and resonances that have mass below $2.5~\mathrm{GeV}/c^2$ as
listed in the Particle Data Group~\cite{ParticleDataGroup:2016lqr}.

As expected, the inclusion of the strange quark in (2+1)-flavor case
leads to a systematic enhancement of both $s/T^3$ and $P/T^4$ throughout
the deconfined phase. This behavior originates from the increase in the
effective number of partonic degrees of freedom. In the two-flavor scenario,
only the light $u$ and $d$ quarks with masses $m_u=m_d= 3.503$ MeV$/c^2$
contribute to the thermodynamics, whereas the $(2+1)$-flavor system includes
an additional strange-quark degree of freedom with mass $m_s=96.4$ MeV$/c^2$, 
thereby increasing the quark degeneracy factor from 6 to 9, in Eq.~(\ref{eflavor}). 
The larger degeneracy results in a higher entropy and pressure, reflecting the
increased number of thermally accessible microscopic states.

The extrapolated HRG results beyond $T_c$, shown by the magenta curves, 
exhibit the characteristic behavior of confined hadronic matter. 
At low temperatures, only the lightest hadronic states are present, 
leading to relatively small values of $s/T^3$ and $P/T^4$. 
As the temperature approaches $T_C$, the contributions
of heavier resonances give rise to a steep increase in both quantities,
signaling the onset of deconfinement and the transition towards partonic
matter~\cite{Tan:2019zyw}.
For comparison, the ideal parton gas results are also shown (blue dashed
curves). Owing to the absence of interactions, the normalized thermodynamic
quantities rapidly approach their Stefan--Boltzmann limits and remain nearly
temperature independent.
The crossover equation of state smoothly interpolates between the HRG and
QGP limits, reproducing the continuous evolution of the thermodynamic
observables expected from lattice QCD across the crossover region.
Since the $(2+1)$-flavor equation of state provides the most realistic
description of QCD matter, all subsequent calculations presented in this
work are performed using the $u$, $d$, and $s$ quark sector~\cite{Tan:2019zyw}.

\begin{figure*}[htbp]
    \centering
    \includegraphics[width=0.8\linewidth]{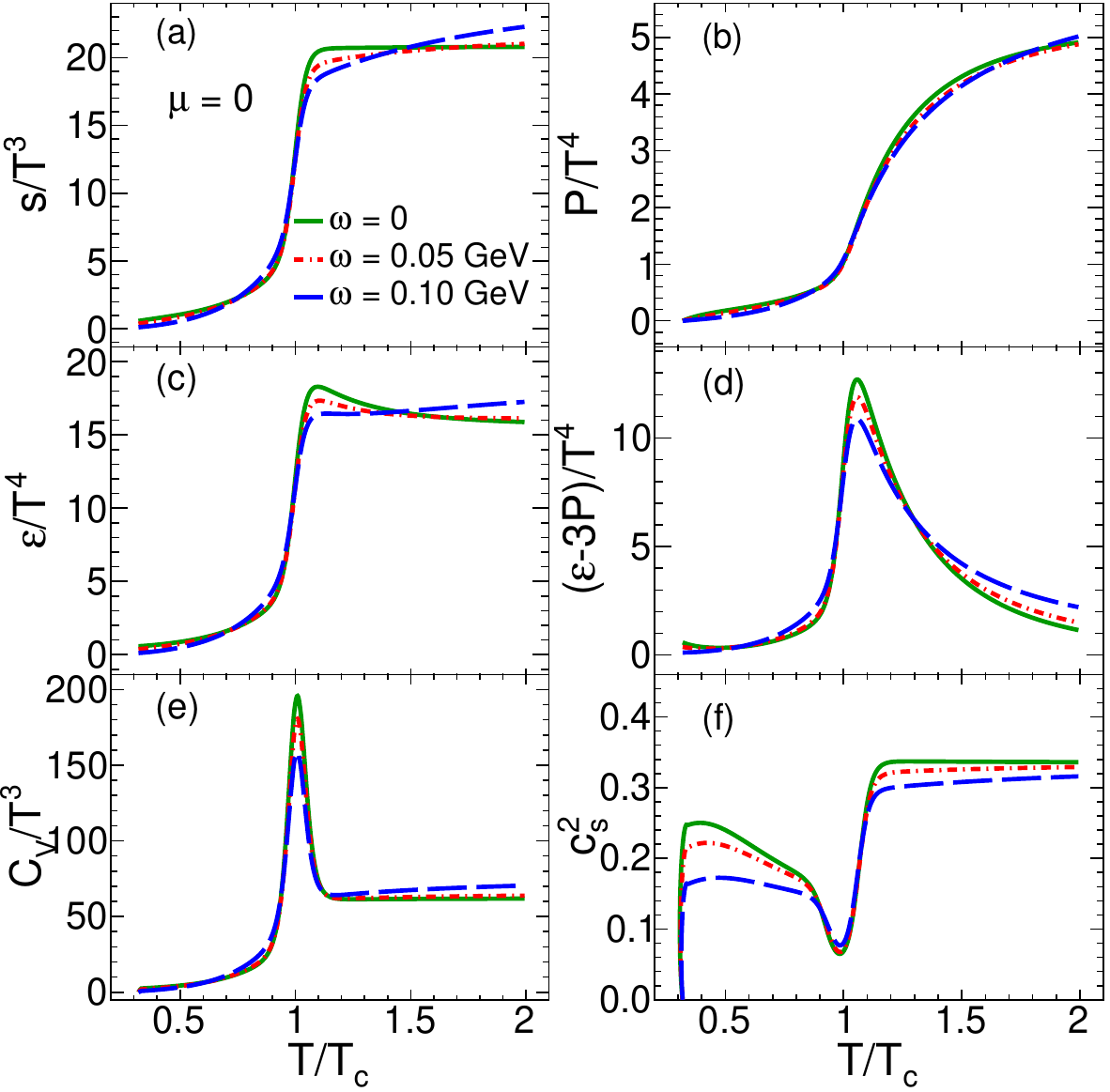}
    \caption{Normalized thermodynamic quantities, $s/T^3$, $P/T^4$, 
    $\varepsilon/T^4$, $\Delta = (\varepsilon-3P)/T^4$, $C_V/T^3$, and $c_s^2$, 
    as a function of $T/T_c$ corresponding to different $\omega$ values ($0$, $0.05$, 
    and $0.10$ GeV) at $\mu=0$.}
    \label{fig3}
\end{figure*}

\subsection{Effect of rotation at zero chemical potential}
Lattice-QCD calculations are most reliable at vanishing chemical
potentials ($\mu_B=\mu_Q=\mu_S=0$), corresponding approximately to the
thermodynamic conditions achieved in the highest-energy heavy-ion collisions
at RHIC ($\sqrt{s_{NN}}=200$ GeV) and LHC energies ($\sqrt{s_{NN}}=2.76$ and
$5.02$ TeV).

In the non-central collisions at such high energies, large
orbital angular momentum is generated, giving rise to a rapidly rotating QGP with
finite vorticity. Motivated by such observation, we investigate the
influence of rotation on the thermodynamic properties of QCD matter at
$\mu = 0$ for three different angular velocities, $\omega=0$, $0.05$,
and $0.10$ GeV. The width of the crossover temperature ($\Gamma$) is
set to be 0.05$T_c$.

Figure~\ref{fig3} shows the dimensionless thermodynamic quantities, 
entropy density ($s/T^3$), pressure ($P/T^4$), energy density
($\varepsilon/T^4$), the trace anomaly ($\Delta = (\varepsilon-3P)/T^4$), 
specific heat at constant volume ($C_V/T^3$), and squared speed of sound 
($c_s^2$), as a function of $T/T_c$ at zero chemical potential. The green 
solid line represent the ideal case, without any rotation, while the red 
and blue dashed line correspond to the cases including finite
rotation $0.05$ GeV and $0.10$ GeV, respectively. 

The entropy density $s/T^3$ exhibits the characteristic crossover 
behavior expected from lattice QCD. At lower temperatures in the HG region, $s/T^3$
approaches asymptotically to zero because only the lightest hadronic
states are produced, while the heavier resonances are exponentially
suppressed by Boltzmann factor. 
It experiences a sharp surge within a 
narrow critical window of width $\Gamma\sim$ $8$ MeV around the transition 
temperature, $T_c$.
Above $T_c$, there is a sharp rise in $s/T^3$, which saturates at the
Stefan-Boltzmann (SB) limit at very high temperature.
Rotation suppresses the entropy in both phases since it modifies the particle 
spectrum and changes the thermal population.
The higher value of rotation, $\omega=0.10$ GeV, increases the 
value of entropy density, at $T/T_c > 1.5$.

A similar trend is observed for $P/T^4$ with rotation.
The smoother curve comes from the 
fact that $P(T)$ is obtained from the relation given by Eq.~\ref{equP},
which is expressed as an integral of entropy density $s$, thus such a
continuous and slow rise is expected.
The non-monotonic behavior in $s/T^3$ and $P/T^4$, above $T_c$ is driven
by the combined effects of the vorticity and the contributions from both
the phases governed by the weighting function $f(T)$.

\begin{figure*}[htbp]
    \centering
    \includegraphics[width=0.8\linewidth]{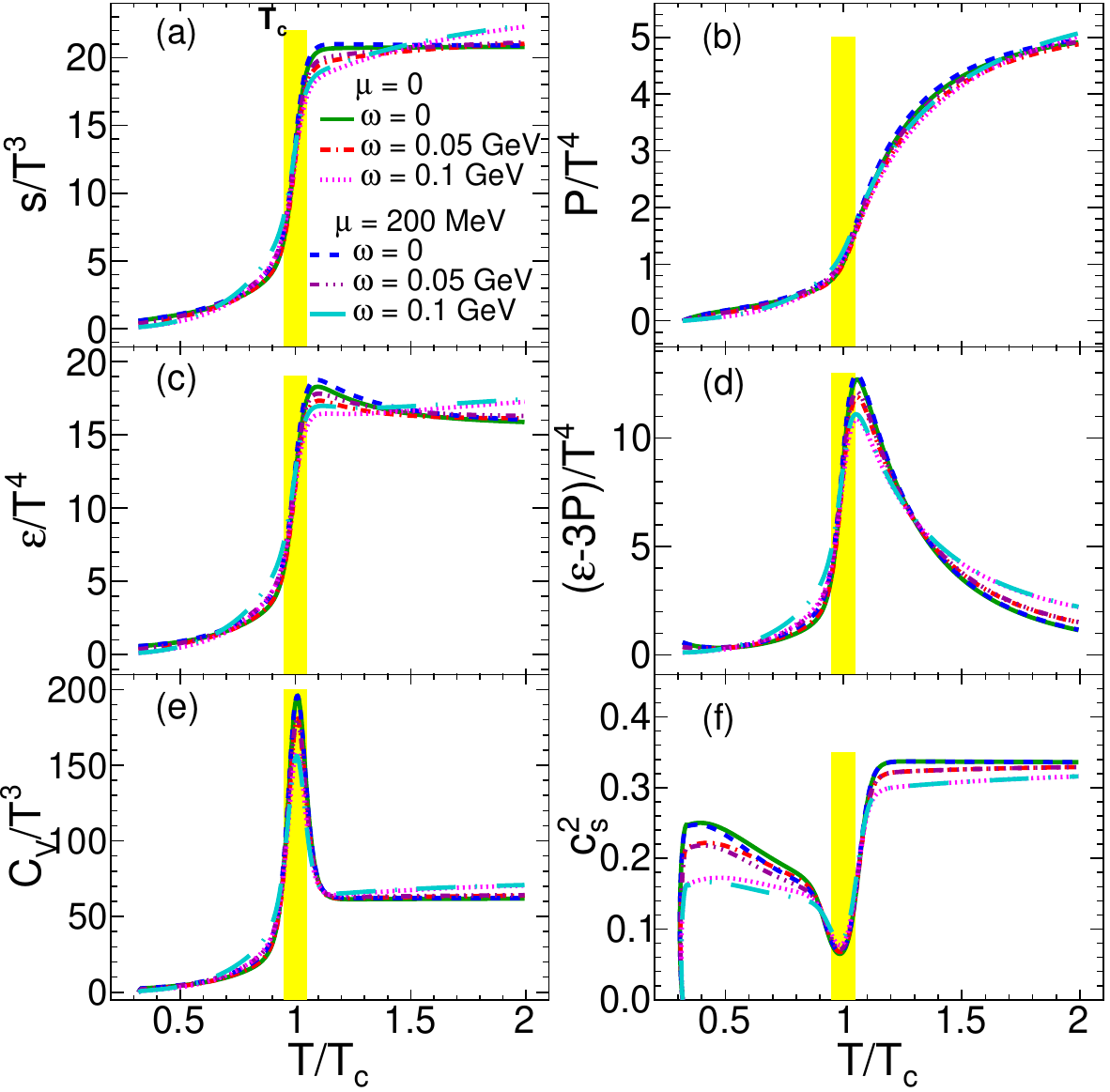}
    \caption{Normalized thermodynamic quantities, $s/T^3$, $P/T^4$, 
    $\varepsilon/T^3$, $\Delta=(\varepsilon-3P)/T^4$, $C_V/T^3$, and $c_s^2$, 
    as a function of $T/T_c$ at vanishing and finite chemical potential, under 
    the effect of different values of rotation ($0$, $0.05$, and $0.10$ GeV). 
    The yellow band corresponds to the crossover region.} 
    \label{fig4}
\end{figure*}

The normalized energy density $\varepsilon/T^4$ shows non-monotonic 
behavior and has a peak just above $T_c$, as shown in Fig.~\ref{fig3} 
for the ideal case. It rises slowly with 
temperature in the HG region, then show a sharp rise and peak near
$T_c$ in the crossover region, and remains constant in the partonic region. 
The peak emerges because the entropy density $(s(T))$ grows drastically just above $T_c$,
while the pressure $(P(T))$ lags behind. The horizontal spread of this peak is determined
by how slowly the pressure $(P/T^4)$ evolves.
%In the hadron gas range, there is negligible effect of rotation. 
The effect of rotation is more prominent in the QGP region. 
The higher value of rotation, particularly $\omega=0.10$ GeV, increases the 
value of energy density, at $T/T_c > 1.5$. 

Trace anomaly, $\Delta$ = ($\varepsilon-3P$)/$T^4$ $-$ the so 
called ‘‘interaction measure’’, provides a measure of the deviation of 
QCD matter from the conformal limit. As expected, $\Delta$ exhibits a 
sharp peak just above $T_c$, as the quarks and gluon degrees of freedom
are set free during deconfinement. At both lower 
temperature HG region and higher temperature partonic region, $\Delta$
asymptotically approaches zero. The presence of rotation suppresses
the height of the peak, indicating that rotational effects soften the
interaction measures in the crossover region. 

The specific heat at constant volume $C_V$ has also a sharp peak near $T_c$,
arising from the rapid variation of the energy density with temperature
[Eq.~(\ref{equCv})]. 
The peak, in the crossover region, reflects the rapid change in degrees of freedom.
The peak reduces in the presence of rotation, 
where as in the partonic phase, rotation increases the value of $C_V$.

The temperature dependence of the square of velocity of sound $c_s^2$
is shown in Fig.~\ref{fig3}. There is a gradual decrease of $C_s^2$
with temperature. As the system approaches crossover, $c_s^2$ decreases
sharply in a narrow temperature region ($T_c-\Gamma$ to $T_c+\Gamma$) and exhibits a
pronounced minimum near $T_c$, signaling the softening of the equation of state
and the slowing of the speed of sound. Unlike energy density and pressure, which 
are determined by the integral of entropy density and show a broad peak or slow rise near $T_c$, 
the speed of sound depends on its temperature derivative and hence shows a sudden drop near $T_c$, 
making it particularly sensitive to crossover dynamics.
The inclusion of rotation further reduces $c_s^2$ over the entire temperature range,
indicating that global rotation softens the equation of state.

\subsection{Effect of rotation at finite chemical potential}
The evolution of the QCD matter with temperature was further extended
to investigate the combined effect of finite baryon chemical potential
and rotation on the QCD equation of state.
We consider the chemical potential of $\mu_B\sim 200$ MeV, corresponding
approximately to the thermodynamic conditions achieved in $Au+Au$ collisions
at $\sqrt{s_{_{NN}}} = 19.6$ GeV. 
Figure~\ref{fig4} shows the temperature dependence of the normalized
entropy density ($s/T^3$), pressure ($P/T^4$), energy density
($\varepsilon/T^4$), interaction measure ($(\varepsilon-3P)/T^4$),
specific heat ($C_V/T^3$), and the squared speed of sound ($c_s^2$) at
both vanishing and finite chemical potential, with and without rotation.

\begin{figure}[t]
    \centering
    \includegraphics[width=\linewidth]{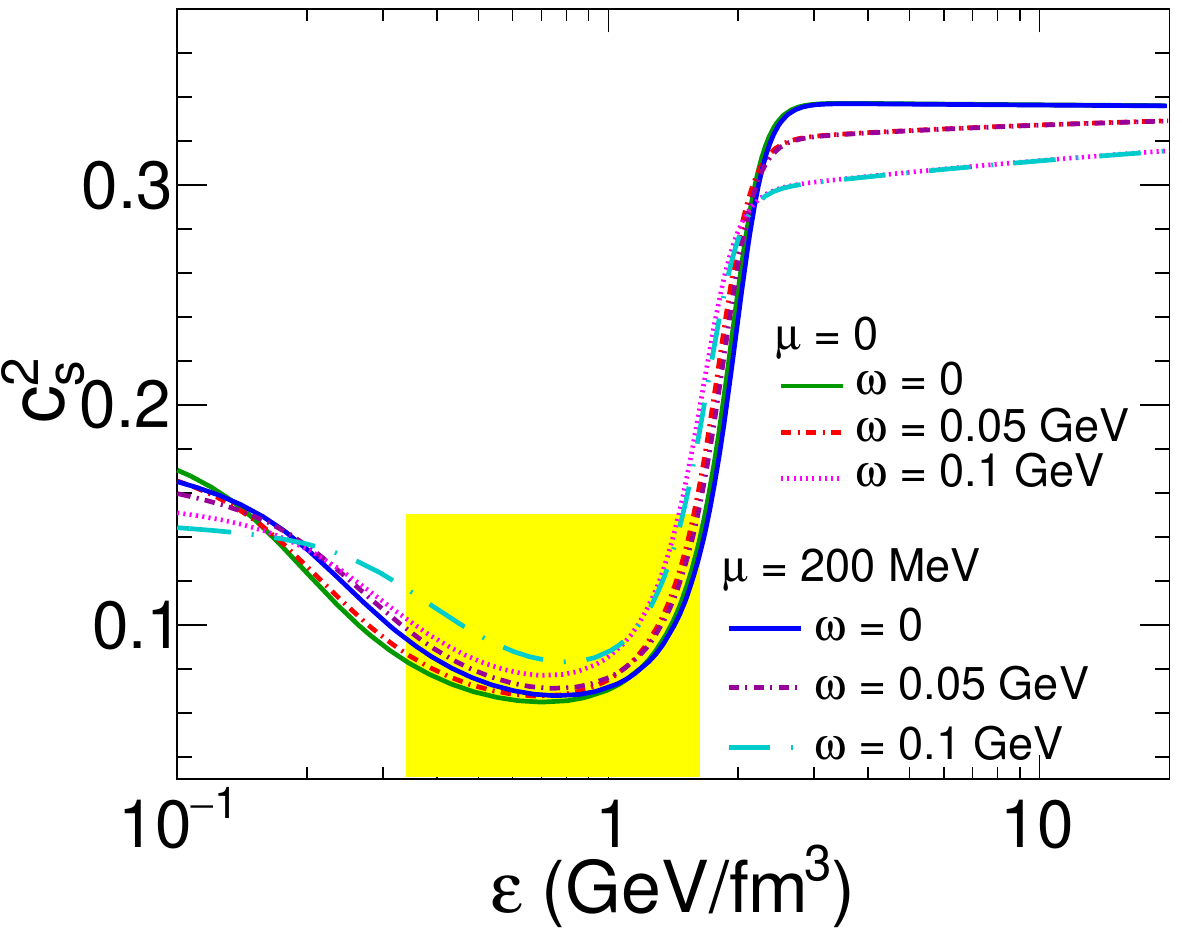}
    \caption{$c_s^2$ as a function of energy density, $\varepsilon$, at 
    vanishing and finite baryon chemical potential, at different rotation 
    ($0$, $0.05$, $0.10$ GeV). The yellow band corresponds to the 
    crossover region.}
    \label{fig5}
\end{figure}

\begin{figure*}[thbp]
    \centering
    \includegraphics[width=0.78\linewidth]{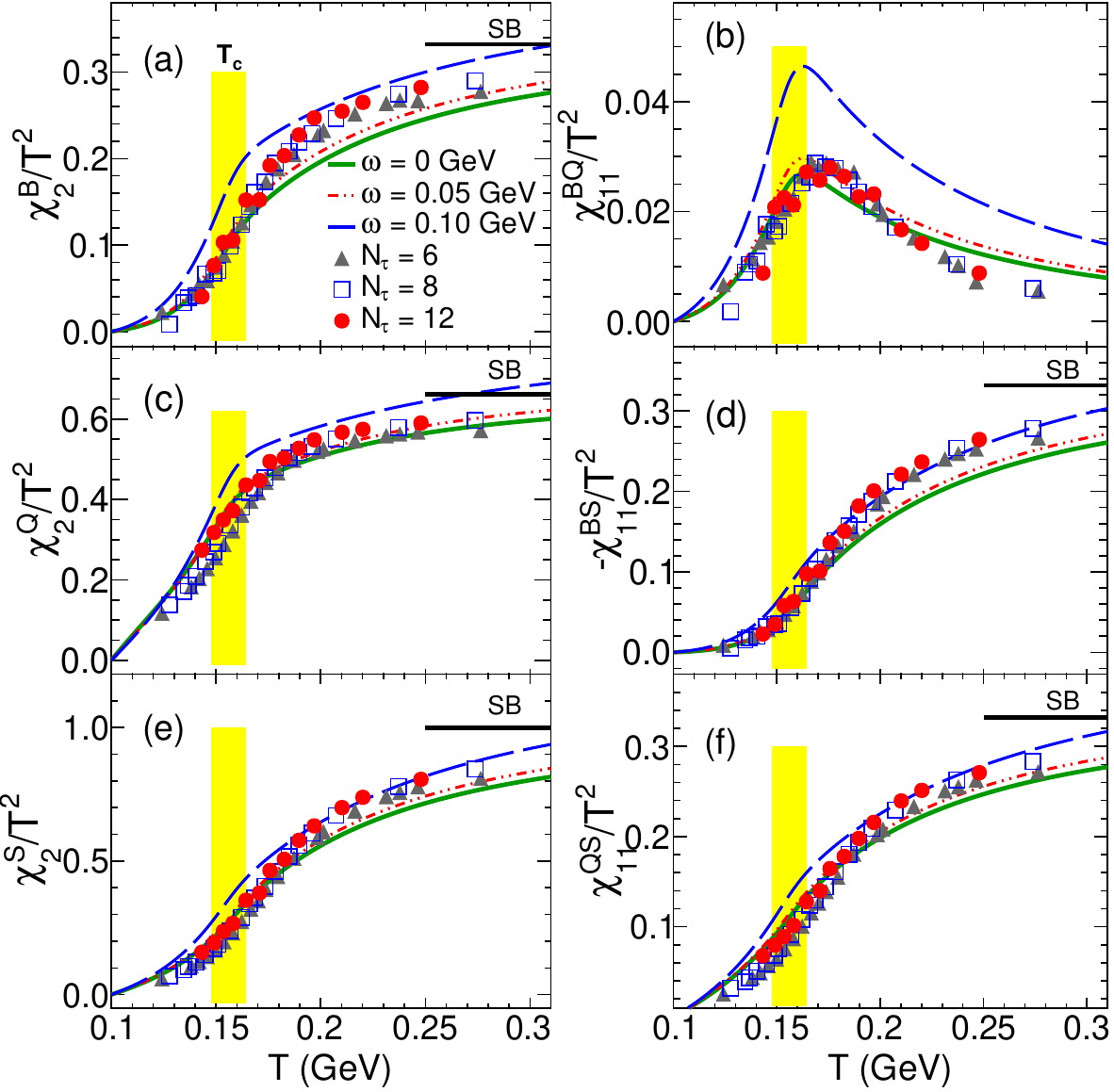}
    \caption{Normalized quadratic susceptibilities ($\chi_2^B/T^2$, $\chi_2^Q/T^2$, 
    and $\chi_2^S/T^2$) and their correlations ($\chi_{11}^{BQ}/T^2$, 
    $-\chi_{11}^{BS}/T^2$, and $\chi_{11}^{QS}/T^2$) as a function of temperature 
    at different rotation ($\omega=0$, $0.05$ GeV, and $0.10$ GeV), at $\mu=0$. The 
    yellow band corresponds to the crossover region. The lattice data points are taken 
    from ref.~\cite{HotQCD:2012fhj}.}
    \label{fig6}
\end{figure*}

The thermodynamic observables exhibit only a modest sensitivity to
non-zero baryon chemical potential. The effects are primarily confined
to the hadronic phase, where a non-zero $\mu_B$ enhances the thermal
population of baryons and strange hadrons, leading to a slight increase
in the bulk thermodynamic quantities. As the system approaches the crossover,
the influence of chemical potential becomes progressively weaker, and above
$T_c$, the normalized thermodynamic observables are nearly insensitive
to $\mu_B=200$ MeV. This behavior reflects the dominance of thermal
excitations over baryon-density effects in the deconfined phase for
moderate chemical potentials relevant to the Beam Energy Scan program at RHIC.
In hadronic phase, there is a slight reduction in $c_s^2$ at
finite baryon chemical potential
within the temperature interval
$0.5 \lesssim T/T_c \lesssim 0.8$, while it remains essentially
unchanged throughout the crossover and deconfined regions.
The reduction of $c_s^2$ at low temperatures arises from the increased
abundance of baryonic and strange degrees of freedom, which soften the
equation of state and reduce the average propagation speed of sound
in the medium.

The softening of the equation of state plays a crucial role in the
hydrodynamic evolution of the fireball produced in relativistic
heavy-ion collisions. Near the minimum of $c_s^2$, commonly referred
to as the softest point of the equation of state, the pressure gradients
are significantly reduced, resulting in the slowing of the 
expansion and cooling of matter in the temperature and energy density
range corresponding to the softest point of the 
equation of state~\cite{HotQCD:2014kol}.
The system spends a longer time in this region, hence
one expects to observe characteristic signatures from this regime.

Figure~\ref{fig5} shows the variation of the square of the speed of sound, 
$c_s^2$, with energy density, $\varepsilon$, to facilitate a direct comparison 
with hydrodynamic simulations and the experimental measurements.
The softest-point structure is clearly visible over the crossover interval,
corresponding to an energy density of approximately
$\varepsilon_c \simeq 0.34$--$1.63~\mathrm{GeV/fm^3}$.
At lower energy densities ($\varepsilon < 1$ GeV/fm$^3$), $c_s^2$ exhibits
a modest enhancement with
chemical potential, whereas there is no effect of chemical potential on
$c_s^2$ in the deconfined phase as a function of $\varepsilon$.
The squared speed of sound increases with rotation up to the energy 
density corresponding to the critical temperature, 
while in higher energy densities it decreases with rotation.
These results demonstrate that, for moderate baryon densities relevant
to intermediate-energy heavy-ion collisions, rotational effects produce a
more pronounced modification of the equation of state than finite
chemical potential.

\subsection{Fluctuations and correlations of conserved charges}
Fluctuations and correlations of conserved charges provide some of
the most sensitive probes of the microscopic structure of strongly
interacting matter and play a central role in exploring the
QCD phase diagram~\cite{Asakawa:2000wh,Jeon:2000wg}.
The generalized susceptibilities are directly
related to quadratic fluctuations of conserved baryon number ($B$),
electric-charge ($Q$), and strangeness ($S$), and therefore encode
information on the active degrees of freedom across the quark-hadron
transition. Their behavior changes markedly across the crossover,
reflecting the transition from hadronic phase to deconfined
quarks and gluons. These observables are accessible both in
lattice QCD calculations and experimentally measured event-by-event
fluctuations of conserved numbers in relativistic heavy-ion collisions,
which provide an important connection between theory and experiment.

Figure~\ref{fig6} shows the temperature dependence of quadratic fluctuations
of conserved charges, $\chi_2^B$, $\chi_2^Q$, and $\chi_2^S$, together
with their correlations, $\chi_{11}^{BQ}$, $\chi_{11}^{BS}$,
and $\chi_{11}^{QS}$, for different rotation values
($\omega$ = $0$, $0.05$ and $0.1$ GeV) at vanishing chemical potential. The
results of our work are compared with the corresponding available non-rotating results 
from lattice QCD calculations~\cite{HotQCD:2012fhj}.

The quadratic fluctuations which are second order susceptibilities of conserved
charges increase rapidly with temperature and exhibit a systematic
enhancement with increasing rotation. This behavior originates from the
growing number of thermally accessible charge-carrying degrees of freedom
as the system evolves from a hadronic medium to a deconfined
quark-gluon plasma. 
In hadronic phase, the dominant carriers are a mixture of particles of
very different masses. The electric charge fluctuations $(\chi_2^Q)$ are governed
primarily by the light charged pions and therefore increase most rapidly
at low temperatures, whereas baryon-number fluctuations $(\chi_2^B)$ and
strangeness fluctuations $(\chi_2^S)$ remain comparatively suppressed because they are dominated by the
substantially heavier nucleons and kaons, respectively.
Near $T_c$, the liberation of quark degrees of freedom in the
deconfined phase produces a pronounced increase in all three 
susceptibilities. As the temperature increases further, the effective
quark masses gradually diminishes in the partonic phase, and the
susceptibilities approach their SB limits which is the characteristic of an ideal
$(2+1)$-flavor quark gas. The enhancement produced by rotation becomes
increasingly pronounced in the deconfined phase, indicating that rotation
favors the thermal excitation of conserved-charge carriers for all temperature ranges.

The mixed susceptibilities provide complementary information on the
correlations among conserved quantum numbers. The temperature dependence
of $\chi_{11}^{BQ}$ for different rotation values is shown in
Fig.~\ref{fig6} (b). The correlations between
baryon number and electric charge $\chi_{11}^{BQ}$, exhibits a
characteristic non-monotonic temperature dependence with a maximum
in the crossover region. At low temperatures, this observable is
dominated by the contribution from charged baryons such as the
proton and $\Delta^{++}$ resonances. In the high temperature limit,
however, the quark masses become effectively massless and weighted sum of
up, down and strange quark charges vanishes, which leads to the vanishing
$\chi_{11}^{BQ}$.  Rotation enhances the magnitude of this correlation
over the entire temperature range.
In case of $\chi_{11}^{BQ}$, which comprises 2$\chi_2^u - \chi_2^d -\chi_2^s$,
the lighter quark susceptibilities approach unity faster than the strange
quark. Therefore,  $\chi_{11}^{BQ}$ value decreases with increase in temperature
after going through the crossover region.

The correlations involving strangeness with baryon number ($\chi_{11}^{BS}$)
and electric charge ($\chi_{11}^{QS}$) are particularly sensitive to the
changes in the strangeness degrees of freedom. The temperature
dependence of $\chi_{11}^{BS}$ and $\chi_{11}^{QS}$ for different
rotation values are shown in Fig..~\ref{fig6} (d) and (f), respectively.
In the hadronic phase, $\chi_{11}^{BS}$ is dominated by strange baryons
such as $\Lambda$, $\Sigma$, $\Xi$, and $\Omega$, which simultaneously
carry baryon number and strangeness. Since these quantum numbers possess
opposite signs, the resulting correlation is negative. In contrast,
$\chi_{11}^{QS}$ receives its principal contributions from charged
strange hadrons, including kaons, $\Sigma^{\pm}$, $\Xi^{\pm}$, producing
a positive correlation between electric-charge and strangeness.
As the temperature approaches $T_c$, the increasing thermal excitation
of strange hadrons and the subsequent liberation of strange quarks lead
to a rapid growth of both correlations. In the partonic phase, with further
increase in temperature, thermal energies substantially exceed the
strange quark mass leading to abundantly populate the strange quark and
antiquark pair. Therefore, both $\chi_{11}^{QS}$ and $\chi_{11}^{BS}$
continue to increase toward their SB limit of three flavor
quark gas. Rotation systematically
enhances the magnitude of both correlations, with the largest effects
observed in the partonic phase.
In case of all the quadratic fluctuations and their correlations, the effect
of rotation is more at higher temperature.
The agreement between our crossover model and non-rotating lattice QCD 
is quite satisfactory over a broad temperature range. It provides a theoretical 
baseline for comparison with possible rotating lattice QCD results.

\begin{figure}[htbp]
    \centering
    \includegraphics[width=\linewidth]{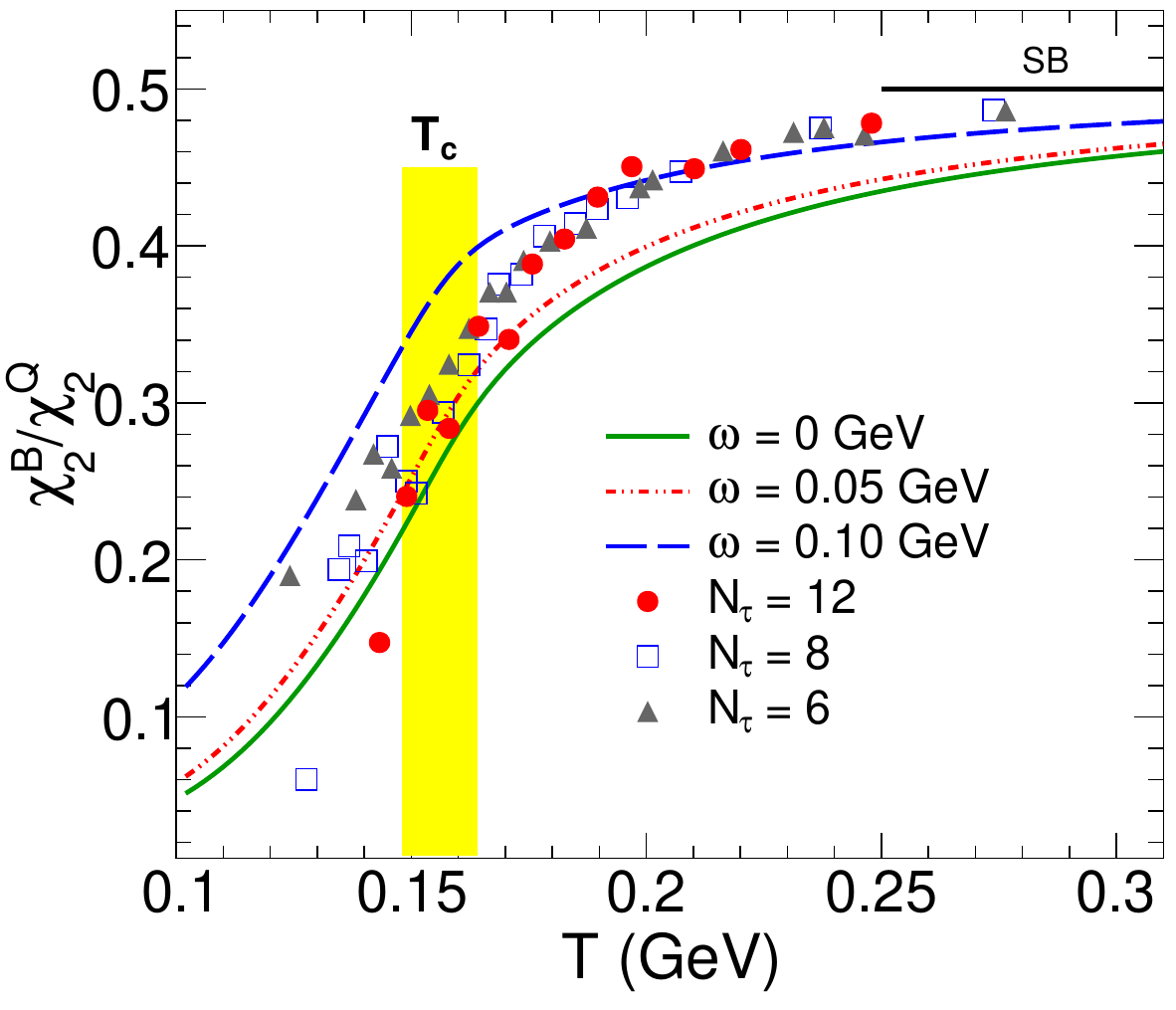}
    \caption{$\chi_2^B/\chi_2^Q$ as a function of temperature at zero chemical
    potential under different rotation ($\omega=0$, $0.05$ GeV, and $0.10$ 
GeV). The yellow band corresponds to the 
    crossover region.}
    \label{fig7}
\end{figure}

To facilitate direct comparison with experimental measurements,
the ratio of net baryon number and electric charge fluctuations $\chi_2^B/\chi_2^Q$ 
is studied, which is closely related to the experimentally
measured ratio of proton to net-charge fluctuations in heavy-ion collisions. 
Figure~\ref{fig7} shows the temperature dependence of $\chi_2^B/\chi_2^Q$ at 
zero chemical potential, along with the results from lattice QCD calculations.
This ratio increases monotonically with rotation and gradually approaches
its ideal-gas (SB) limit at high temperatures.

\begin{figure}[ht]
    \centering
    \includegraphics[width=\linewidth]{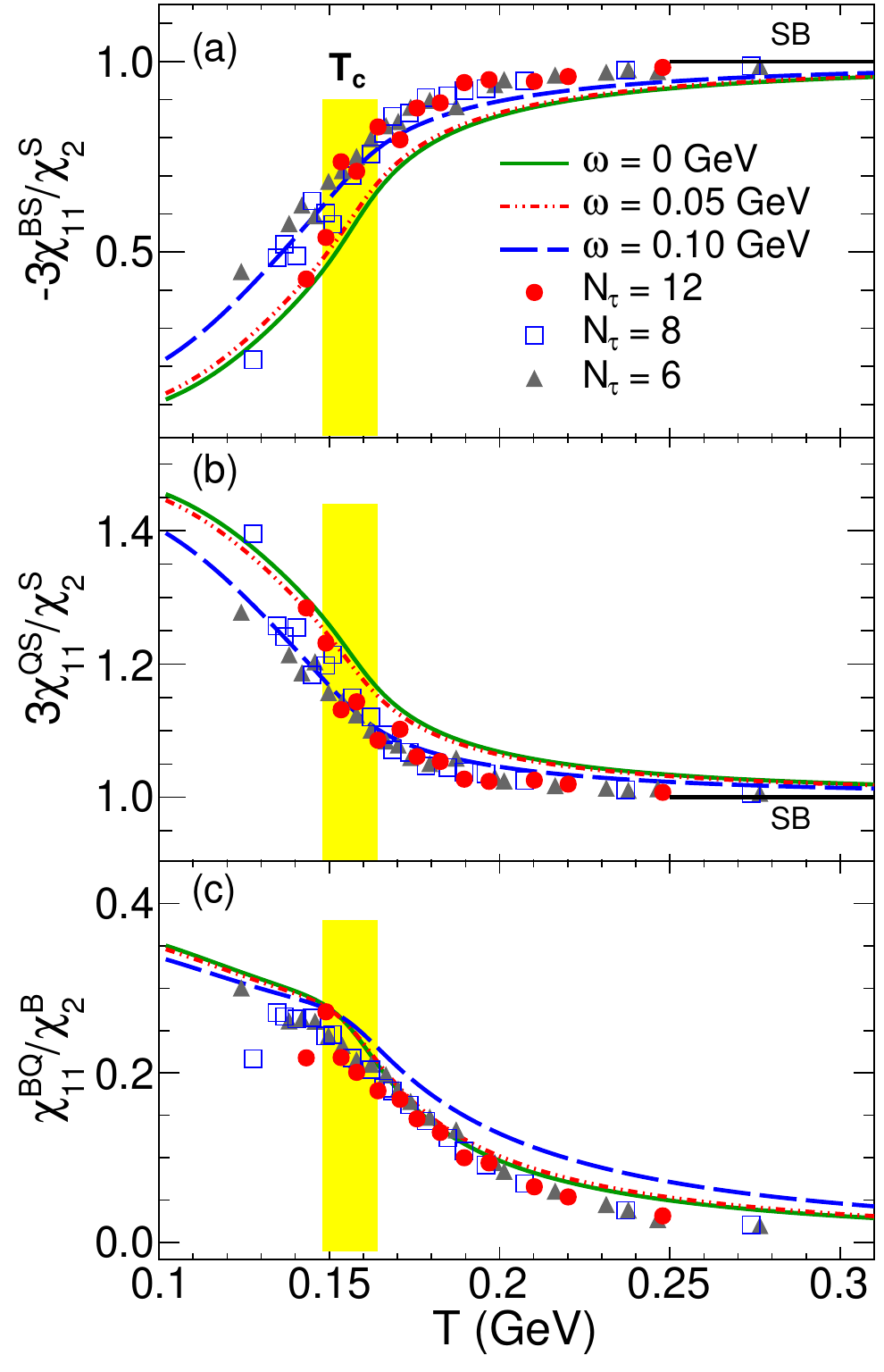}
    \caption{$-3\chi_{11}^{BS}/\chi^S_2$, $3\chi_{11}^{QS}/\chi^S_2$, and
    $\chi_{11}^{BQ}/\chi^B_2$ as a function of temperature at zero chemical 
potential under different rotation ($\omega=0$, $0.05$ GeV, and $0.10$ GeV). The yellow band corresponds to the 
    crossover region.}
    \label{fig8}
\end{figure}

Further insight into the relevant charge carriers is obtained from
the temperature dependence of the normalized correlation ratios,
$-3\chi_{11}^{BS}/\chi^S_2$, $3\chi_{11}^{QS}/\chi^S_2$, and 
$\chi_{11}^{BQ}/\chi^B_2$, under the effect of different rotation
values 0, 0.05 and 0.1 GeV as shown in Fig.~\ref{fig8},
which are compared with the corresponding available non-rotating results 
from lattice QCD calculations~\cite{HotQCD:2012fhj}.
By forming these suitable combination of ratios, the leading order corrections
can be removed in the lattice calculations allowing more stringent
comparisons between theoretical predictions, lattice-QCD calculations,
and experimental data from heavy-ion collision. The quantities $-3\chi_{11}^{BS}/\chi_2^S$,
$3\chi_{11}^{QS}/\chi_2^S$, and $\chi_{11}^{BQ}/\chi_2^B$ provide robust
signatures of the underlying microscopic degrees of freedom.
At high-temperature limit, both $-3\chi_{11}^{BS}/\chi_2^S$ and
$3\chi_{11}^{QS}/\chi_2^S$ approach unity, consistent with the expectations
for an ideal gas of weakly interacting quarks~\cite{Koch:2005vg}, whereas 
$\chi_{11}^{BQ}/\chi_2^B$ shows a different behavior as leading order
perturbative corrections do not cancel completely~\cite{HotQCD:2012fhj}.
The ratios $-3\chi_{11}^{BS}/\chi^S_2$ and $\chi_{11}^{BQ}/\chi^B_2$ increases with
rotation, on the other-hand $3\chi_{11}^{QS}/\chi^S_2$ decreases with
rotation. These results demonstrate that global rotation not only
modifies the magnitude of conserved-charge fluctuations but also alters
their mutual correlations, highlighting the important role of vorticity
in shaping the thermodynamic response of QCD matter across the crossover.

\section{Conclusion}\label{conclusion}
In this study, we have developed a thermodynamically consistent hybrid
equation of state to study the QCD crossover region based on a smooth
interpolation between hadron gas and parton gas under the influence
of global rotation. The proposed framework provides a unified description
of the QCD crossover region and enables a systematic investigation of
the combined effects of rotation and finite baryon chemical potential
on the thermodynamic properties of QCD matter.

The model successfully reproduces the characteristic crossover behavior
predicted by lattice QCD and captures the continuous evolution of the
equation of state from the confined hadronic phase to the deconfined
partonic phase. We find that the inclusion of strange quarks significantly
enhances the thermodynamic observables owing to the increased number of
partonic degrees of freedom, demonstrating the important role of flavor
in determining the equation of state at high temperature.

This study reveals that, 
the thermodynamic quantities such as pressure, energy density, entropy
density, trace anomaly, and specific heat
are enhanced with increasing rotation at higher temperature, and suppressed 
in the crossover region, whereas the squared speed of sound is suppressed 
over the entire temperature range.. This indicates that rotation softens 
the equation of state and modifies the thermodynamic response of strongly 
interacting matter. For finite baryon chemical potential ($\mu_B=200$ MeV), 
results show that the effects of moderate baryon density remain relatively small
compared with those induced by rotation. The influence of finite $\mu_B$
is largely confined to the hadronic phase, while the thermodynamic
properties of the deconfined medium remain nearly unchanged.

We have further investigated the fluctuations and correlations of
conserved charges under the influence of rotation.
The present model successfully reproduces the temperature dependence
of the quadratic fluctuations of the conserved numbers
($\chi_2^B$, $\chi_2^Q$ and $\chi^S_2$)
and their correlations ( $\chi_{11}^{BQ}$,  $\chi_{11}^{BS}$ and
$\chi_{11}^{QS}$). These quantities
exhibit a systematic enhancement with increasing rotation, with the
rotational effects becoming more pronounced in deconfined partonic phase.

The present study demonstrates that global rotation constitutes an
important thermodynamic control parameter that can significantly
influence both the equation of state and the fluctuation observables
of QCD matter. Since conserved-charge fluctuations are directly
accessible through event-by-event measurements in relativistic
heavy-ion collisions, the predicted rotational modifications
may provide experimentally testable signatures of vortical QCD
matter produced in non-central collisions at RHIC and the LHC. The
hybrid equation of state developed in this work therefore offers a
useful framework for incorporating rotational effects into hydrodynamic
and transport simulations and for interpreting future precision
measurements of the QCD phase structure.

%\begin{acknowledgments}
%\vspace{-0.2cm}
%\end{acknowledgments}

\nocite{*}
\bibliographystyle{apsrev4-1}
\bibliography{crossover}     % Produces the bibliography via BibTeX.

\end{document}